\documentclass[twocolumn,preprintnumbers,amsmath,amssymb]{revtex4}

\usepackage{graphicx}
\usepackage{dcolumn}
\usepackage{bm}
\usepackage[utf8x]{inputenc}
\usepackage{float}
\usepackage{tikz}
\begin{document}


\title{A Study of Excited $\Omega_b^-$ States in Hypercentral Constituent Quark Model via Artificial Neural Network}

\author{Halil Mutuk}
 \email{halilmutuk@gmail.com}
\affiliation{Physics Department, Faculty of Arts and Sciences, Ondokuz Mayis University, 55139, Samsun, Turkey}%


\begin{abstract}
In this work, we have obtained mass spectra, radiative decay widths and strong decay widths of newly observed excited $\Omega_b^-$ states, i.e. $\Omega_b(6316)^-$, $\Omega_b(6330)^-$, $\Omega_b(6340)^-$, and $\Omega_b(6350)^-$. Mass spectrum is obtained in Hypercentral Constituent Quark Model (hCQM) by solving six-dimensional nonrelativistic Schrödinger equation via Artificila Neural Network (ANN). In this respect, radiative decay widhts are calculated by a generalization of a framework from meson to baryon. Also, strong decay widths of the low-lying $\Omega_b$ states within $^3P_0$ model are calculated. Obtained results are presented with the comparison of available experimental data and other theoretical studies. 
\end{abstract}

\maketitle

\section{\label{sec:level1}Introduction}
In the past years, significant experimental progress has been made in heavy baryon physics. In resemblance with the atomic spectroscopy, where transitions of the lines in the spectrum gives information about the interaction between the atomic core and the electrons, energy of the resonances gives information about interaction inside the nucleons. According to quak model, hadrons are strongly interacting particles that are made of quarks. 

In this context, the heavy baryons have been got much attention theoretically and experimentally in the last decade. Single heavy baryon system $(Qqq)$ is composed of a heavy $(Q)$ quark and two light quarks $(qq)$ which is bound via gluon interactions. This system is the QCD analogue of the helium atom in which two electrons (light particles) are orbitting around a fixed proton (heavy particle) bounded by the electromagnetic interactions. 

Very recently, the LHCb collaboration reported four narrow peaks in the $\Xi_b^0 K^-$ mass spectrum \cite{1}
\begin{eqnarray}
\Omega_b(6316)^- : M&=& 6315.64 \pm 0.31 \pm 0.07 \pm 0.50  ~ \text{MeV}, \nonumber \\ 
\Gamma &<& 2.8  ~ \text{MeV} \\
\Omega_b(6330)^- : M&=& 6330.30 \pm 0.28 \pm 0.07 \pm 0.50  ~ \text{MeV}, \nonumber \\ 
\Gamma &<& 3.1  ~ \text{MeV} \\
\Omega_b(6340)^- : M&=& 6339.71 \pm 0.26 \pm 0.05 \pm 0.50  ~ \text{MeV}, \nonumber \\ 
\Gamma &<& 1.5  ~ \text{MeV} \\
\Omega_b(6350)^- : M&=& 6349.88 \pm 0.35 \pm 0.07 \pm 0.50  ~ \text{MeV}, \nonumber \\ 
\Gamma &=&1.4^{+1.0}_{-0.8} \pm 0.1   ~ \text{MeV},
\end{eqnarray}
where the uncertainties are statistical, systematic and the last is due to the knowledge of $\Xi_b^0$ mass. The significances of these states are $2.1 \sigma$ for $\Omega_b(6316)^-$, $2.6 \sigma$ and exceed $5 \sigma$ for  $\Omega_b(6340)^-$ and $\Omega_b(6350)^-$. 

Before the LHCb's measurement there is a vast literature for the mass spectrum of the excited $\Omega_b$ states by various quark models \cite{2,3,4,5,6,7}, QCD Sum Rules \cite{8,9,10,11,12} and Regge phenomenology \cite{13,14,15}. The experimental result has already stimulated theoretical works. In Ref. \cite{16}, Chen et al. systematically studied the internal structure of $P$-wave $\Omega_b$ baryons, and calculated their mass spectra and strong decay properties by QCD Sum Rules method. They concluded that  all the four excited $\Omega_b$ baryons recently discovered by LHCb can be well explained as $P$-wave $\Omega_b$ baryons. In Ref. \cite{17}, Liang and Lü studied the strong decays of the low-lying $\Omega_b$ states within the $^3P_0$ model and pointed out that these excited states can be reasonably assigned as $\lambda$-mode $\Omega_b(1P)$ states. The possible interpretations of the four peaks as the meson-baryon molecular stateswere were studied by Liang and Oset \cite{18}. They claimed four states at higher energies which are
seen as peaks with the same statistical significance as the first of the
states claimed. Wang used explicit $P$-wave between the two $s$-quarks to construct thecurrent operators to study the $P$-wave $\Omega_b$ states with the full QCD sum rules by carrying out the operator product expansion up to the vacuum condensates of dimension 10 \cite{19}. Xiao et al. elaborated OZI allowed two-body strong decays of the low-lying $\lambda$-mode $\Omega_b$ baryons up to N=2 shell using the chiral quark model within the $j-j$ coupling scheme \cite{20}.

As a result of the studies mentioned above, these newly observed excited $\Omega_b$ states can be viewed as $P$-wave states. In quark models, the $\Omega_b$ states have three valence quarks, $ssb$. If we don't include an additional $P$-wave to this system, we can obtain the ground states of $\Omega_b$ as $\Omega_b(J^P=\frac{1}{2}^+)$ and $\Omega_b(J^P=\frac{3}{2}^+)$. For the time being, only the $\Omega_b(\frac{1}{2}^+)$ is observed \cite{21}. But if there exist a relative $P$-wave between two $s$-quarks or between $ss$ diquark and $b$ quark, seven negative parity $\Omega_b$ states can be obtained. Furthermore, if this relative $P$-wave excitation takes about an energy of 300-500 MeV, then $P$-wave $\Omega_b$ states have the mass in the range of 6300-6400 MeV. Quark models and diquark-quark models finds similar mass values \cite{22,23}.

In the present work, we will study $\Omega_b^-$ states in quark model via artificial nerual network. In Section \ref{sec:level2}, Hypercentral Constituent Quark Model (hCQM) is presented. Section \ref{sec:level3} is devoted to the Artificial Neural Network and application of it to quantum mechanical systems. In Section \ref{sec:level4}, we present our results for mass spectrum, radiative and strong decay widths comparing with available experimental data and theoretical studies. A short summary is concluded in the last section.

\section{\label{sec:level2}Quark Model of $\Omega_b^-$ State}
A baryon is an example of quantum mechanical version of classical three-body system. Three-quark systems with a heavy quark $(Q)$ of mass $m_Q$ and two light quarks $(qq)$ of equal mass $m_q$ can be studied in this prescribed way. Single heavy baryons with one heavy quark $(Q)$ ($c$ or $b$) and two light quarks ($u$, $d$ and $s$) provide a laboratory for studying the dynamics of the light quarks in the presence of a heavy quark. Two states which are in the same mode but have different spin numbers may mix in the heavy quark limit since light-quark spin-spin force is still on work in this limit.

In this present work, we use non-relativistic framework of hypercentral Constituent Quark Model (hCQM) which is well established for the study of the properties of  baryons \cite{24,25,26,27,28}.

\subsection{Hypercentral Constituent Quark Model (hCQM)}
The fundamental idea of the hypercentral approach to three-body systems is simple. The coordinates of the bodies in the system are rewritten in terms of the relative coordinates. In this case, two relative coordinates $(\vec{\rho}, \vec{\lambda})$ are rewritten into a  non-relativistic Schrödinger equation in the six-dimensional space. The hyrperradius $x$ and hyperangle $\xi$ is given as

\begin{equation}
x= \sqrt{\vec{\rho}^2 + \vec{\lambda^2}}, ~ \xi=\arctan \frac{\rho}{\lambda}.
\end{equation}

To remove the center of mass motion of the system, Jacobi coordinates can be used \cite{26}
\begin{eqnarray}
\vec{\rho}&=&\frac{1}{\sqrt{2}}(\vec{r}_1-\vec{r}_2), \\
\vec{\lambda} &=& \frac{m_1 \vec{r}_1+m_2 \vec{r}_2-(m_1+m_2)\vec{r}_3}{m_1^2+m_2^2+(m_1+m_2)^2},
\end{eqnarray}
where $r_i$ and $m_i=(1,2,3)$ denote the spatial and the mass coordinate of the $i$−th constitute quark. The reduced mass corresponding to Jacobi coordinates $\vec{\rho}$ and $\vec{\lambda}$ are

\begin{eqnarray}
m_{\rho}&=&\frac{2m_1m_2}{m_1 + m_2}, \\
m_{\lambda}&=&\frac{2m_3(m_1^2 + m_2^2 + m_1m_2)}{(m_1+m_2)(m_1+m_2+m_3)},
\end{eqnarray}
where $m_i=(1,2,3)$ are the constituent quark masses. The angles of the hyperspherical coordinates are given by $\Omega_{\rho}=(\theta_{\rho}, \phi_{\rho} )$ and $\Omega_{\rho}=(\theta_{\lambda}, \phi_{\lambda} )$. 

After having removed the center of mass motion by using Jacobi coordinates, the kinetic energy operator can be written as
\begin{eqnarray}
\frac{P_x^2}{2m}&=&-\frac{\hbar^2}{2m}(\Delta_{\rho} + \Delta_{\lambda}) \nonumber \\ &=& -\frac{\hbar^2}{2m} \left( \frac{\partial^2}{\partial x^2} + \frac{5}{x} \frac{\partial}{\partial x} + \frac{L^2(\Omega)}{x^2}  \right)
\end{eqnarray}
where $m=\frac{2m_{\rho}m_{\lambda}}{m_{\rho}+m_{\lambda}}$ is the reduced mass and $L^2(\Omega)=L^2 (\Omega_{\rho}, \Omega_{\lambda}, \xi)$ is the six dimensional generalization of the squared angular momentum operator and  is a representation of quadratic Casimir operator of the six-dimensional rotational group $O(6)$. Its eigenfunctions are the hyperspherical harmonics $Y_{[\gamma] l_{\rho} l_{\lambda}} (\Omega_{\rho}, \Omega_{\lambda}, \xi)$ satisfying the eigenvalue relation
\begin{equation}
L^2 Y_{[\gamma] l_{\rho} l_{\lambda}} (\Omega_{\rho}, \Omega_{\lambda}, \xi) =-\gamma (\gamma + 4) Y_{[\gamma] l_{\rho} l_{\lambda}} (\Omega_{\rho}, \Omega_{\lambda}, \xi).
\end{equation}

Here $\vec{L}=\vec{L}_{\rho} + \vec{L}_{\lambda}$, $l_{\rho}$ and $ l_{\lambda}$ are the angular momenta associated with the Jacobi coordinates $(\vec{\rho}, \vec{\lambda})$ and $\gamma$ is the hyperangular momentum quantum number satisfying $\gamma=2n +  l_{\rho} + l_{\lambda} $ where $n$ is a nonnegative integer. Notice that in the hCQM, it is hard to seperate $\rho-$ mode excitations and $\lambda-$ mode excitations since the sets ($l_{\rho}=1, l_{\lambda}=0$) which is $\rho-$ mode excitation and ($l_{\rho}=0, l_{\lambda}=1$) which is $\lambda-$ mode exciation give the same hyperangular momentum quantum number, $\gamma$.  



The potential for the hCQM can be written in terms of hyperradius $(x)$ as
\begin{equation}
\sum_{i < j} V(r_{ij})=V(x) + \cdots.
\end{equation}
In this approximation, the three-quark potential is a function of only the hyperradius $x$ and the dependence on the single pair coordinates cannot be disentangled from the third one. Therefore it contains not only  two-body interactions but also contains three-body interactions. The case of three-body interactions are important in the study of hadrons since the existence of a direct gluon-gluon interaction-which is the non-abelian feature of QCD- produce three-body forces. 

The potential is a key ingredient in the study of meson and baryon spectra. In the present study, we consider the hypercentral potential $V(x)$ as the hyper Coulomb plus linear potential with second order correction  and spin-dependent interaction, which is given
as follows \cite{29}:
\begin{equation}
V(x)=V^0(x)+\left( \frac{1}{m_{\rho}} + \frac{1}{m_{\lambda}} \right) V^1(x)+V^2(x)+V_{SD}(x),
\end{equation}
where $V^0(x)$ is given by
\begin{equation}
V^0(x)=\frac{\tau}{x} + \beta x,
\end{equation}
where $\tau=-\frac{2}{3}\alpha_s$ is the hyper Coulomb strength corresponding to the strong running coupling constant $\alpha_s$ and $\beta$ is the string tension of the confinement part. 

The first-order correction reads as:
\begin{equation}
V^{(1)}(x)=-C_F C_A \frac{\alpha_s^2}{4x^2},
\end{equation}
where, $C_F$ and $C_A$ are the casimir charges of the fundamental and adjoint representation and   $\alpha_s$ is the strong running coupling constant.

The second-order correction is given by  (with the notation of $m=\frac{2m_{\rho}m_{\lambda}}{m_{\rho}+m_{\lambda}}$) \cite{30}:
\begin{eqnarray}
V^2(x)&=&-\frac{C_F D_{1,s}^{(2)}}{2m^2} \left\lbrace \frac{1}{x}, \vec{p}^2 \right\rbrace + \frac{C_F D_{2,s}^{(2)}}{2m^2}\frac{1}{x^3}\vec{L}^2 \nonumber \\ &+& 
\frac{\pi C_F D_{d,s}^{(2)}}{m^2} \delta^{(3)}(\vec{x}) + \frac{4\pi C_F D_{S^2,s}^{(2)}}{3m^2} \vec{S}^2 \delta^{(3)}(\vec{x}) \nonumber \\ &+& \frac{3 C_F D_{LS,s}^{(2)}}{2m^2} \frac{1}{x^3} \vec{L} \cdot \vec{S} \nonumber \\  &+& \frac{C_F D_{S_{12},s}^{(2)}}{4m^2}\frac{1}{x^3}S_{12} (\hat{x}),
\end{eqnarray} 
where $S_{12}(\vec{r})\equiv 3\hat{x} \cdot \vec{\sigma}_1 \hat{x} \cdot \vec{\sigma}_2 - \vec{\sigma}_1-\vec{\sigma}_2$ and $\vec{S}=\vec{\sigma}_1 /2 + \vec{\sigma}_2 / 2$. 

In order to obtain the expressions for the coefficients, one has to perform the matching between nonrelativistic QCD (NRQCD) and potential nonrelativistic QCD (pNRQCD). A procedure for this mathcing can be found in \cite{31,32}. After matching, one can obtain the relevant contributions as follows \cite{30}:

\begin{eqnarray}
\delta D_{1,s}^{(2)} &=& \alpha_s(x), \nonumber \\
\delta D_{2,s}^{(2)} &=&  \alpha_s(x), \nonumber \\
\delta D_{d,s}^{(2)} &\simeq & \alpha_s(x) \left\lbrace 1 + \frac{\alpha_s}{\pi} \left( \frac{2C_F}{3} + \frac{17C_A}{3} \right) \ln mx \right\rbrace, \nonumber \\
\delta D_{S^2,s}^{(2)} &\simeq & \alpha_s(x) \left\lbrace 1- \frac{7C_A}{4} \frac{\alpha_s}{\pi} \ln mx \right\rbrace,  \nonumber \\
\delta D_{LS,s}^{(2)} &\simeq & \alpha_s(x) \left\lbrace 1- \frac{2C_A}{3} \frac{\alpha_s}{\pi} \ln mx \right\rbrace,  \nonumber \\
\delta D_{S_{12},s}^{(2)} &\simeq & \alpha_s(x) \left\lbrace 1- C_A \frac{\alpha_s}{\pi} \ln mx \right\rbrace.
\end{eqnarray}

One interesting point here is that, spin-dependent potentials do not have $\ln mx$ contributions. The spin-dependent part of the potential reads as follows:
\begin{eqnarray}
V_{SD}(x)&=& V_{SS} (\vec{S}_{\rho} \cdot \vec{S}_{\rho}) + V_{\gamma S}(x)(\vec{\gamma} \cdot \vec{S}) \nonumber \\ &+& V_T(x) \left[ S^2 - \frac{3 (\vec{S} \cdot \vec{x}) (\vec{S} \cdot \vec{x})}{x^2} \right],
\end{eqnarray}
where, $\vec{S}=\vec{S}_{\rho}+\vec{S}_{\rho}$ with $\vec{S}_{\rho}$ is the spin vector of $\vec{\rho}$ and  $\vec{S}_{\lambda}$ is the spin vector of $\vec{\lambda}$. 

The spin dependent potential contains three types of the interaction: the spin–spin term $V_{SS}$ which gives spin singlet triplet splittings, the spin–orbit term $V_{\gamma S}$ and tensor term $V_T$ give the fine structure of the states \cite{33}. Defining $V_V(x)=\frac{\tau}{x}$ as vector part and $V_S(x)=\beta x$ as scalar part of the static potential, these spin-dependet terms can be written as follows: 
\begin{eqnarray}
V_{SS}(x)&=& \frac{1}{3 m_{\rho} m_{\lambda}} \nabla^2 V_V, \\
V_{\gamma S}(x)&=&\frac{1}{2 m_{\rho} m_{\lambda}x} \left( 3\frac{dV_v}{dx}- \frac{dV_s}{dx} \right),\\
V_T(x) &=&  \frac{1}{6 m_{\rho} m_{\lambda}} \left( 3\frac{d^2V_v}{dx^2}- \frac{1}{x}\frac{dV_v}{dx} \right). 
\end{eqnarray}

The Hamiltonian of the three-quark system in the hQCM can be written as
\begin{equation}
H=\frac{P_x^2}{2m}  + V(x),
\end{equation}
where $m=\frac{2m_{\rho}m_{\lambda}}{m_{\rho}+m_{\lambda}}$ is the reduced mass and $x$ is the six-dimensional radial hypercentral coordinate of the three-quark system. After writing the Hamiltonian the next step is solving Schrödinger equation, $H \Psi= E \Psi$. 

In conventional quark model, Schrödinger equation is solved in three-dimensional space. In hCQM, Schrödinger equation lives in six-dimensional space and needs to be handled. This can be done in conventional way such as numerically. In this work, we have solved six-dimensional Schrödinger equation by using Artificial Neural Networks, (ANNs). There are some advantages of using ANN for solving differential equations \cite{34}
\begin{itemize}
\item Computational complexity does not increase quickly in the ANN method when the number of sampling points is increased while in the other standard numerical methods computational complexity increases rapidly as  the number of sampling points increased in the interval
\item Model based on neural network offers an opportunity to tackle in real time difficult differential equation problems arising in many sciences and engineering applications.
\item The ANN method to solve a differential equation provides a solution with very good generalization properties.
\end{itemize}

The main advantage of using ANN is that computational complexity does not increase considerably when the number of dimensions in problem increase. The other point is that the  solution  is  continuous  overall the domain of integration.

\section{\label{sec:level3}Formalism of Artificial Neural Network}
Especially the last decade has seen a significant raise of interest in machine learning where the learning part is done by the artificial neural networks. Artificial neural networks (ANNs) are systems of simple processing elements (called "neurons") managing their data and communicating with other elements.

\begin{figure}[H]
\includegraphics[width=3.4in]{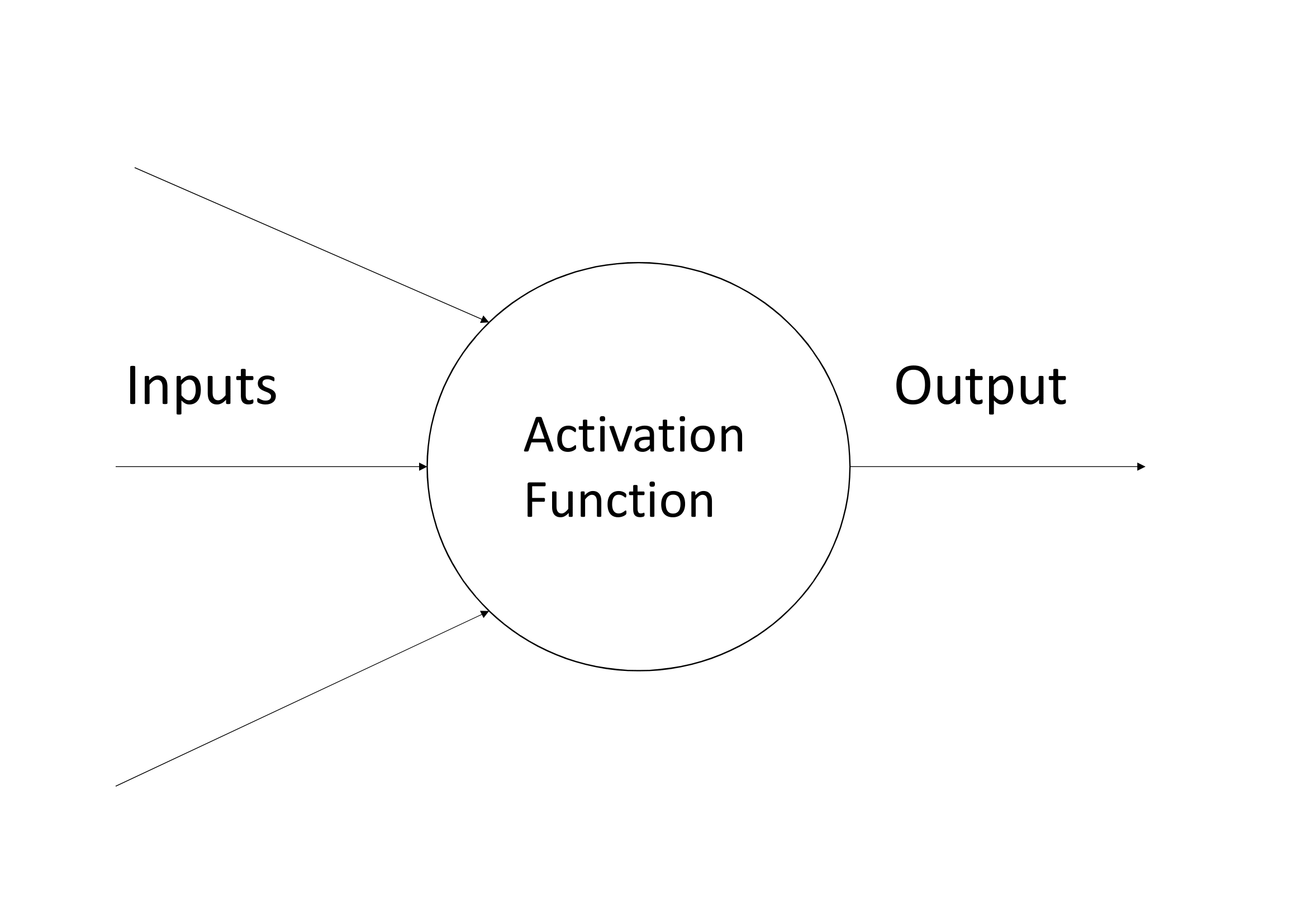}
\caption{\label{fig:2} A model of single neuron}
\end{figure}
\begin{figure}[H]
\includegraphics[width=3.4in]{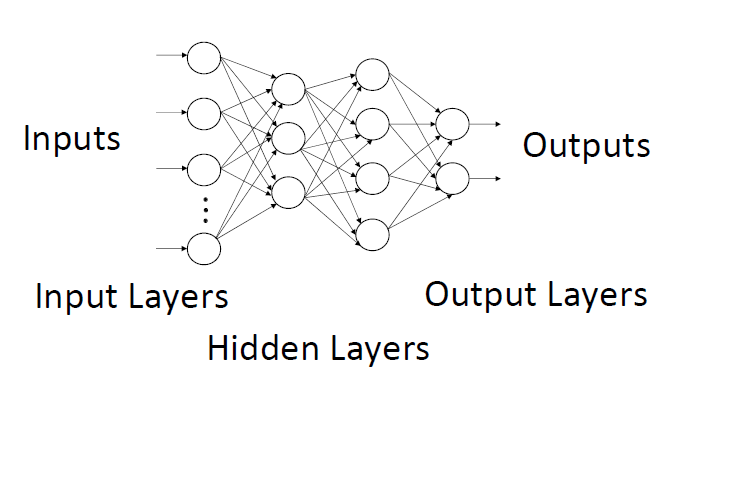}
\caption{\label{fig:5} Multilayer neural network}
\end{figure}

The fundamental ingredient of an artificial neural network is neuron (perceptron in computerized systems). Figure \ref{fig:2} represents a single artificial neuron.
Each neuron receives any number of input and produces only one output. If this output comes from input layers, it will be an input for the hidden layers. In this manner, the inputs are the outputs of activation functions in where the inputs are multiplied by the connection weights. This activation function (neuron transfer function) determines the output. In practice, one single neuron is not capable of handling problems. That's why networks composed of neurons are being used. In Figure \ref{fig:5}, the architecture of a multilayer perceptron is shown.

The most common architecture of ANNs is the multilayer feed forward network.   In this study, we consider a feed forward neural network with one input layer, one hidden layer and one output layer. The signals are propagated from the input layer to the output layer where each processing element is responsible for operating the signals coming from preceding layer and sending information to the connected next layer. Fig. \ref{fig:2} is an example of feed forward neural network.



\subsection{Mathematical Model of an Artificial Neural Network}
The relationship of the input and output of the nertwork layers can be written as follows:
\begin{eqnarray}
o_i&=&\sigma(n_i), \\
o_j&=&\sigma(n_j), \\
o_k&=&\sigma(n_k),
\end{eqnarray}
where $i$ is for input, $j$ is for hidden and $k$ is for output layers. Input to the neurons read as
\begin{eqnarray}
n_i&=&(\text{Input signal to the NN}), \\
n_j&=& \sum_{i=1}^{N_i} \omega_{ij}o_i+ \theta_j, \\
n_k&=& \sum_{i=1}^{N_j} \omega_{jk}o_j+ \theta_k,
\end{eqnarray}
where, $N_i$ and $N_j$ are the numbers of units in the input and hidden layers, $\omega_{ij}$ is the synaptic weight parameter which connects the neurons $i$ and $j$ and $\theta_j$ and $\theta_k$ are the threshold parameters of the neurons $j$ and $k$, respectively \cite{35}. The output of the network defined as 
\begin{equation}
o_k=\sum_{j=1}^{b_n} \omega_{jk}\sigma \left( \sum_{i=1}^{a_n}\omega_{ij}n_i +\theta_j \right)+ \theta_k. \label{eqn1}
\end{equation}
One needs derivative of above function since it will be used in evaluation of error function. This can be obtained as
\begin{eqnarray}
\frac{\partial o_k}{\partial \omega_{ij} }&=& \omega_{jk} \sigma^{(1)}(n_j)n_i,\\
\frac{\partial o_k}{\partial \omega_{jk} }&=& \sigma(n_j)\delta_{kk^\prime},\\
\frac{\partial o_k}{\partial \theta_j }&=& \omega_{jk} \sigma^{(1)}(n_j),\\
\frac{\partial o_k}{\partial \theta_k }&=&\delta_{kk^\prime}.
\end{eqnarray}

Activation function of a neuron defines the output of that neuron for  given an input or inputs. In this work, sigmoid function
\begin{equation}
\sigma(x)=\frac{1}{1+e^{-x}} \label{sigmoid}
\end{equation}
is used as an activation function.  This function is continuous in everywhere and belongs to class $C^\infty$. Therefore, it is possible to derive all the derivatives of $\sigma(x)$. This is an important aspect for the Schrödinger equation since it is a second order differential equation. 

\subsection{Application to Quantum Mechanics}
Consider the following differential equation
\begin{equation}
H\Psi(x)=f(x). \label{eqn2}
\end{equation}
Here $H$ is a linear operator, $f(x)$ is a known function and  $\Psi(x)=0$ at the boundaries. A general method for soving this equation is to try a function which is believed to be a solution. For this reason, a trial function such as
\begin{equation}
\Psi_t(\textbf{x})=A(\textbf{x})+B(\textbf{x}, \textbf{$\eta$})N(\textbf{x}, \textbf{p}) \label{eqnart}
\end{equation}
can be written. This trial function depends on $\textbf{p}$ and $\textbf{$\eta$}$ which are to be neural network parameters. $\textbf{p}$ is the weight and bias of the artificial neural network. In Eq. \ref{eqnart}, $A(\textbf{x})$ and $B(\textbf{x}, \textbf{$\eta$})$ should be adjusted so that $\Psi_t(\textbf{x})$ satisfies boundary conditions of the given equation. In order to solve Eq. (\ref{eqn2}), collocation method can be used. In this method, one choose a finite-dimensional space for trial solutions and a number of points in the domain. After doing that, given differential equation can be transfomed into a minimization problem
\begin{equation}
\underset{p,\eta}{\min} \sum_i \left[ H\Psi_t(x_i)-f(x_i)  \right]^2. \label{dif2}
\end{equation}
In terms of eigenvalue equation, Eqn. (\ref{eqn2}) takes the form
\begin{equation}
H\Psi(x)=\epsilon \Psi(x), 
\end{equation}
where $\Psi(x)$ satisfies the boundary condition, $\Psi(x)=0$. In this case, the trial solution can be written as
\begin{equation}
\Psi_t(x)=B(\textbf{x}, \textbf{$\eta$})N(\textbf{x}, \textbf{p}).
\end{equation}
Here, $B(\textbf{x}, \textbf{$\eta$}) =0$ at boundary conditions for a range of $\eta$ values. If we discretize the domain of the problem, which is a part of collocation method, Schrödinger equation can be transformed into a minimization problem, as mentioned before:
\begin{equation}
E(\textbf{p},\textbf{$\eta$})=\frac{\sum_i \left[ H\Psi_t(x_i, \textbf{p},\textbf{$\eta$})-\epsilon \Psi_t(x_i, \textbf{p},\textbf{$\eta$})  \right]^2}{\int \vert \Psi_t \vert^2 d\textbf{x}}.
\end{equation}
Here $E$ is the error function and $\epsilon$ can be computed as
\begin{equation}
\epsilon=\frac{\int  \Psi_t^{\ast} H \Psi_t  d\textbf{x}}{\int \vert \Psi_t \vert^2 d\textbf{x}}.
\end{equation}
For further details of application of neural networks to quantum mechanical problems, see Ref. \cite{36}.

The three-quark wave function can be factorized as 
\begin{equation}
\psi_{3q}(\vec{\rho}, \vec{\lambda})=\psi_{\gamma \nu}(x)  Y_{[\gamma] l_{\rho} l_{\lambda}} (\Omega_{\rho}, \Omega_{\lambda}, \xi). \label{3quark}
\end{equation}
Here, hyperradial wave function $\psi_{\gamma \nu}(x) $ is labeled by the grand angular quantum number $\gamma$ and by the number of nodes $\nu$. The angular-hyperangular part of the $3q$-state is completely described by the hyperharmonics and is same for any hypercentral potential. The dynamics is contained in the hyperradial wave function \citep{26}: 

\begin{eqnarray}
\psi_{\gamma \nu}(x)&=& \left[  \frac{v! (2g)^6}{(2g+2\nu+5)(\nu+2\gamma+4)!^3} \right]^{1/2} \nonumber \\ &\cdot & (2gx)^{\gamma} e^{-gx} L_{\nu}^{2\gamma +4}(2gx),
\end{eqnarray}
where $g=\frac{\tau m}{\gamma + \nu + 5/2}$. 

 Then, the hyperradial Schrödinger equation can be written as follows:
\begin{equation}
\left[ \frac{d^2}{dx^2} + \frac{5}{x}\frac{d}{dx}-\frac{\gamma(\gamma + 4)}{x^2}     \right]  \psi_{\gamma \nu}(x)= -2m \left[ E-V(x) \right] \psi_{\gamma \nu}(x).
\end{equation}

We used a trial wave function as
\begin{equation}
\psi_t(x)= x^{5/2}\psi_{\gamma \nu}(x) N(x,u,w,v), 
\end{equation}
with $N$ being a feed forward neural network with one hidden layer and $m$ sigmoid hidden units
\begin{equation}
N(x,u,w,v)=\sum_{j=1}^m v_j \sigma(w_jx+u_j).
\end{equation}

To obtain desired results, the first thing that ANN has to do is learning. The learning mechanism is the most important property of ANN. In this work, we used a feed forward neural network with a back propagation algorithm which is also known as delta learning rule. This learning rule is valid for continuous activation function, such as Eqn. \ref{sigmoid}. The algorithm is as follows \cite{37}:
\begin{enumerate}
\item[Step 1] Initialize the weights $w$ from the input layer to the hidden layer
and weights v from the hidden layer to the output layer. Choose the
learning parameter (lies between 0 and 1) and error $E_{max}$. Initially error is taken as 0.
\item[Step 2] Train the network.
\item[Step 3] Compute the error value.
\item[Step 4] Compute the error signal terms of the output layer and the hidden layer.
\item[Step 5] Compute components of error gradient vectors.
\item[Step 6] Check the weights if they are properly modified.
\item[Step 7] If $E=E_{max}$ terminate the training session. If not, go to step 2
with $E \to 0$ and initiate a new training.
\end{enumerate}

The main point to train the network is to initialize the eigenvalue $(E)$ (error function) is to zero and train the network with 150 equidistance points in the interval after selecting one hidden unit. The aim is to get energy function to be zero or at least tends to be zero. If the convergence  is not achieved, then the eigenvalue would be wrong. In that case, eigenvalue should be changed and tried again. If the error function value did not converge to zero after doing these steps, the number of the hidden units must be changed (increased) and run a new cycle. We keep doing this until the error function converges to zero.

By employing this approach it is possible to obtain energy eigenvalues of the Schrödinger equation. We trained the network with some equidistance points in the intervals with $m=5$ and solved the Schrödinger equation in the interval $0 < x < 1$ with $\psi(0)=0$ at the boundaries.  We did not analyze the number of hidden units in the present paper. The optimum value was obtained after doing some calculations.

\section{\label{sec:level4}Numerical Results and Discussion}
The numerical parameters and constants used in this work are

\begin{eqnarray*}
m_s&=& 0.500 ~ \text{GeV},\\
m_b&=& 4.67 ~ \text{GeV}, \\
C_F &=& 2/3, \\
C_A &=& 3 , \\
n_f&=& 3,\\
\alpha_s &=& 0.6  ~ (\mu_0 = \text{1 GeV}), \\
\tau &=& -0.4 \\
\beta &=& 0.18
\end{eqnarray*}

The comparison of the obtained mass spectra with other quark models is shown in Table \ref{tab:table1} with the spectroscopic notation   ($n^{2s+1}L_j$) except $L$ (angular momentum quantum number) is replaced by $\gamma$ (hyper-angular momentum quantum number) in hCQM.

\begin{table*}
\caption{\label{tab:table1}Comparison of mass values of excited $\Omega_b^-$ states with other quark models.  Results are in MeV.}
\begin{ruledtabular}
\begin{tabular}{cccccccc}
State & This Work &  RQM \cite{5} & NQM \cite{6} & hCQM \cite{14} & QM \cite{23}   \\ \hline
$1^{2}P_{1/2}$ & 6314 & 6330 & 6333 & 6338 & 6305 $\pm$ 15\\
$1^{2}P_{3/2}$ & 6330 & 6331 & 6336 & 6328 & 6317 $\pm$ 19\\
$1^{4}P_{1/2}$ & 6339 & 6339 & 6340 & 6343 & 6313 $\pm$ 15\\
$1^{4}P_{3/2}$ & 6342 & 6340 & 6344 & 6333 & 6325 $\pm$ 19\\
$1^{4}P_{5/2}$ & 6352 & 6334 & 6345 & 6320 & 6338 $\pm$ 20\\

\end{tabular}
\end{ruledtabular}
\end{table*}

It can be seen from Table \ref{tab:table1}  that the obtained mass values are in good agreement with the other theoretical studies. The LHCb collaboration observed four narrow structures of excited $\Omega_b^-$ states, namely $\Omega_b^-(6316)$, $\Omega_b^-(6330)$, $\Omega_b^-(6340)$ and $\Omega_b^-(6350)$ in the $\Xi_b^0 K^-$ mass spectrum. The theory predicts five states whereas experiment observed only four states. The reamining state should have a broad width which can hardly be observed in experiments. The quantum numbers of newly observed excited $\Omega_b^-$ states are not available. Some recent efforts are done and the results can be seen in Table 2.

\begin{table}[H]
\caption{\label{tab:table2} $J^{PC}$ quantum number assignments to the excited $\Omega_b^-$ states of recent studies.}
\begin{ruledtabular}
\begin{tabular}{cccccc}
State & \cite{16} & \cite{17} & \cite{19} \\ \hline

$\Omega_b^-(6316)$ & $\frac{1}{2}^-$ or $\frac{3}{2}^-$   & $\frac{1}{2}^- $ & $\frac{3}{2}^- $   \\
$\Omega_b^-(6330)$ & $\frac{1}{2}^-$ & $\frac{3}{2}^- $ & $\frac{1}{2}^- $   \\
$\Omega_b^-(6340)$ & $\frac{3}{2}^-$  & $\frac{3}{2}^- $ & $\frac{5}{2}^- $   \\
$\Omega_b^-(6350)$ & $\frac{3}{2}^-$ partner with $\frac{5}{2}^-$ & $\frac{5}{2}^- $ & $\frac{3}{2}^- $   \\
\end{tabular}
\end{ruledtabular}

\end{table}

Since there is no available data upto now on quantum numbers of these excited states, there is no possibility to rule out them.

\subsection{Radiative Decays of Excited $\Omega_b^-$ States}
The study of electromagnetic transitions of  baryons is an important issue for understanding the internal structure of baryons. To calculate the partial widths of the E1 radiative transitions of the $(n ^{2s+1}L_j \to n'^{2s'+1}L'_{j'}+\gamma)$ process, we use \cite{38}:
\begin{eqnarray}
\Gamma_{E_1}& =& \frac{4}{3} C_{fi} \delta_{ss'} e_q^2 \alpha \vert \left\langle R_f \vert r \vert R_i  \right\rangle   \vert ^2 E_\gamma ^3
\end{eqnarray}
where $e_q$ is the quark charge, $\alpha$ is the fine-structure constant, $E_\gamma$ is the final photon energy. The matrix element $C_{fi}$ is given as
\begin{equation}
C_{fi}=\max(L,L^\prime)(2J^\prime+1) \left\lbrace \begin{array}{ccc} L\prime & S^\prime & J  \\ J & L & 1 \end{array} \right\rbrace.
\end{equation}
In order to calculate matrix elements, we need $ \left\langle R_f \vert r \vert R_i  \right\rangle $, where $R$ represents the radial wave function of the meson. In this work, we generalize radial wave function to hyperradial wave function, $r$ to hyper radius $x$, $L$ (angular momentum quantum number) is replaced by $\gamma$ (hyper-angular momentum quantum number), and meson mass to the baryon mass. For the masses of the initial and final states, we used our calculated values. Apart from hCQM, this generalization from meson to baryon electromagnetic transitions were successfully done in \cite{39,40}. The matrix elements of $ \left\langle R_f \vert r \vert R_i  \right\rangle $ were calculated in ANN framework by setting suitable quantum numbers for initial and final states hyperradial wave functions, $\psi_{\gamma \nu}(x)$.  In Table \ref{tab:table3}, we present the partical widths of the radiative decays of excited $\Omega_b^-$ baryons. Our predictions are compatible with the result of Ref. \cite{24}. Radiative decays of $1^{2}P_{1/2}$ and $1^{2}P_{3/2}$ to  $\Omega_b \gamma$ and $1^{4}P_{3/2}$ and $1^{4}P_{3/2}$ to $\Omega_b^\ast \gamma$ are very different than others. If these decays are studied in further experiments, this situation can be identified. 

  \begin{table}[H]
\caption{\label{tab:table3}Partial widths for the radiative decays of the excited $\Omega_b^-$ states. Results are in keV.}
\begin{ruledtabular}
\begin{tabular}{cccccccccc}
 &\multicolumn{2}{c}{ $\Gamma [\Omega_b \gamma]$}&\multicolumn{2}{c}{$\Gamma [\Omega_b^\ast \gamma]$}\\
State &  This work  &  \cite{46} & This work & \cite{46}&  \\ \hline

$1^{2}P_{1/2}$ & 138 & 154 & 1.41 & 1.49 \\
$1^{2}P_{3/2}$& 71.8 & 83.4 & 1.45 & 1.51  \\
$1^{4}P_{1/2}$& 0.85& 0.64 & 104.29 & 99.23  \\
$1^{4}P_{3/2}$& 2.03 & 1.81 & 73.65 & 70.68 \\
$1^{4}P_{5/2}$& 1.02 & 1.21 & 51.57 & 63.26 \\

\end{tabular}
\end{ruledtabular}
\end{table}

\subsection{Decay Widths of Excited $\Omega_b^-$ States }
Besides the mass prediction, the investigation of other physical properties such as radiative decays of the considered resonances would be helpful to identify these resonances more reliably. The study of strong decay of these resonances is also helpful in this respect. The study of the strong decay processes is a challenge task both theoretically and experimentally. 

In this work, we adopt the $^3P_0$ model \cite{41,42,43,44} to calculate  the two-body strong decay widths of the low-lying $\Omega_b$ states. This model is an efficient model to compute open-flavor strong decays of hadrons in the quark model framework. In this model a hadron decay occurs in its rest frame and proceeds via the creation of an additional $q\bar{q}$ pair. This quark-antiquark pair is created with vacuum quantum numbers, $J^{PC}=0^{++}$. The decay widhts can be computed by 
\begin{equation}
\Gamma_{A \to BC}= \frac{2\pi \gamma_0^2}{2 J_A + 1} \Phi_{A \to BC}(q_0) \sum_{M_{J_A},M_{J_B}} \vert \mathcal{M}^{M_{J_A},M_{J_B}} \vert ^2.
\end{equation}
Here, $\vert \mathcal{M}^{M_{J_A},M_{J_B}} \vert$ is the $A \to BC$ amplitude, $\gamma_0$ is the dimensionless pair creation strength, $q_0$ is the relative momentum between $B$ and $C$. The coefficient $\Phi_{A \to BC}(q_0)$ is the phase space factor for the decay. The nonrelativistic expression for this factor reads as
\begin{equation}
\Phi_{A \to BC}(q_0) =2 \pi q_0\frac{M_b M_c}{M_A}.
\end{equation}
The baryon wave function is choosen as $\psi_{3q}(\vec{\rho}, \vec{\lambda})=\psi_{11}(x)  Y_{[1]} l_0 l_1$. The free parameter (pair creation strength) $\gamma_0$ is taken as $9.3$. For the wave function and other issues please see Refs. \citep{23,45}. In the bottom baryon systems, the heavy quark symmetry should be preserved quite well, that is, the light quark spin $\vec j= \vec s_1 + \vec s_2 + \vec L$ is an approximate good quantum number rather than the total spin $\vec S =\vec s_1 + \vec s_2 + \vec s_3$.  The physical resonances should correspond to the mixing of the states in the $L-S$ coupling scheme adopted in this work. The mixing of the states with same $J^P$ is taken into account for the strong decay decay behaviours. The obtained decay widths are given in Table \ref{tab:table4}.

\begin{table}[H]
\caption{\label{tab:table4}Strong decay widths of $\Gamma (\Omega_b \to \Xi_b K^-) $. Results are in MeV.}
\begin{ruledtabular}
\begin{tabular}{cccccccc}
State & This work &  \cite{23} & \cite{46}  \\ \hline
$1^{2}P_{1/2}$ & 0.64 & 0.50 & 49.53 \\
$1^{2}P_{3/2}$ & 2.97 & 2.79 & 1.90 \\
$1^{4}P_{1/2}$ & 1.52 & 1.14 & 95.08 \\
$1^{4}P_{3/2}$ & 0.46 & 0.62 & 0.29 \\
$1^{4}P_{5/2}$ & 2.62 & 4.28 &  1.66 \\

\end{tabular}
\end{ruledtabular}
\end{table}
As can be seen from Table \ref{tab:table4}, our results are  in good agreement with Ref. \cite{23} and compatible with Ref. \cite{46}. The discrepancy in two results for the states of $1^{2}P_{1/2}$ and $1^{2}P_{1/2}$ of Ref. \cite{46} is interesting. Regarding the strong decays for baryon states, no satisfactory model has been achieved. So the reason for this discrepancy may be as a result of non-existence of a single model for explaining baryon open-flavor strong decays.

\section{\label{sec:level5} Summary and Concluding Remarks}

In this paper, we have employed hypercentral constituent quark model to compute mass spectra by solving six-dimensional Schrödinger equation by using artificial neural network. 
The obtained mass spectra are in good agreement with other works. Given the uncertainties of different quark models predictions, these structures are good candidates of the $\Omega_b(1P)$ states. As mentioned before, there are five states for $\Omega_b$ in the quark model. Since there are only four observed $\Omega_b$ states and one is remaining, the status and fate of remaining state is still unclear. 

We also obtained radiative decay  of excited $\Omega_b^-$ states by generalizing the method given in Ref. \cite{38}. The results are appealing for this generalization and we hope this can be used in further studies. 
The radiative decay widths for $\Omega_b^-$ states is in compatible with the reference study. Within the results of radiatie decays, $1^{2}P_{1/2}$ and $1^{2}P_{3/2}$ to $\Omega_b \gamma$ and $1^{4}P_{1/2}$ and $1^{2}P_{3/2}$ to $\Omega_b^\ast \gamma$ are very large than the others.

Strong decay widths are calculated via a generalization of $^3P_0$ model. Obtained results are in good agreement with available experimental data and theoretical studies.  The results for $1^{2}P_{1/2}$ and $1^{4}P_{1/2}$ of Ref. \cite{46} are roughly order of magnitude larger than our results. 

We hope  future experiments on radiative and strong decay widhts of excited $\Omega_b^-$ states will clarify the spin assignments and the status of the  missing state which have not been observed yet. 


\end{document}